\documentclass[11pt]{article}

\usepackage[preprint]{acl}

\usepackage{times}
\usepackage{latexsym}

\usepackage[T1]{fontenc}

\usepackage[utf8]{inputenc}

\usepackage{microtype}

\usepackage{inconsolata}

\usepackage{graphicx}

\title{Topological Structure Learning Should Be A Research Priority for LLM-Based Multi-Agent Systems}

\author{
\textbf{
Jiaxi Yang$^{1}$\thanks{These authors contributed equally to this research.}\hspace{0.5em}, Mengqi Zhang$^{2}$\footnotemark[1]\hspace{0.5em}, Yiqiao Jin$^{3}$\footnotemark[1]\hspace{0.5em}, Hao Chen$^{4}$, Qingsong Wen$^{5}$} \\ 
\textbf{Lu Lin$^{1}$, Yi He$^{2}$, Srijan Kumar$^{3}$, Weijie Xu$^{6}$, James Evans$^{7}$, Jindong Wang$^{2}$\thanks{Corresponding to: Jindong Wang <jwang80@wm.edu>}}
\\
$^1$The Pennsylvania State University \quad $^2$William \& Mary \quad $^3$Georgia Institute of Technology \\
$^4$AMD \quad $^5$Squirrel AI \quad $^6$Amazon \quad $^7$University of Chicago \\
\url{https://TopoAgent.github.io/}
}

\usepackage{algorithm}
\usepackage{algorithmicx}
\usepackage{algpseudocode}
\usepackage{amsmath}
\usepackage{amsthm}
\usepackage{amsfonts}
\usepackage{array}
\usepackage{booktabs}
\usepackage{calc}
\usepackage{colortbl}
\usepackage{enumitem}
\usepackage{footnote}
\usepackage{graphicx}
\usepackage{hyperref}
\usepackage{cleveref}
\usepackage{listings}
\usepackage{mathtools}
\usepackage{microtype}
\usepackage{multirow}
\usepackage{newfloat}
\usepackage{seqsplit}
\usepackage{subcaption}  
\usepackage{url}
\usepackage{xcolor}
\usepackage{xspace}
\usepackage{lineno}
\usepackage{pifont}
\usepackage{wrapfig}
\usepackage{cleveref}
\usepackage{lipsum}

\newcommand{\eg}{{\em e.g.,~}} 
\newcommand{\ie}{{\em i.e.,~}}

\newtheoremstyle{openproblemstyle}
  {\topsep}
  {\topsep}
  {\itshape}
  {}
  {\bfseries}
  {.}
  { }
  {\thmname{#1}~\thmnumber{#2}}

\theoremstyle{openproblemstyle}
\newtheorem{openproblem}{Open Problem}

\newtheoremstyle{researchdirectionstyle}
  {\topsep}
  {\topsep}
  {\itshape}
  {}
  {\bfseries}
  {.}
  { }
  {\thmname{#1}~\thmnumber{#2}}

\theoremstyle{researchdirectionstyle}
\newtheorem{researchdirection}{Research direction}

\newtheoremstyle{positionstyle}
  {\topsep}
  {\topsep}
  {\itshape}
  {}
  {\bfseries}
  {.}
  { }
  {\thmname{#1}~\thmnumber{#2}}

\theoremstyle{positionstyle}
\newtheorem{position}{P}

\begin{document}
\maketitle

\begin{abstract}
Large Language Model-based Multi-Agent Systems (MASs) have emerged as a powerful paradigm for tackling complex tasks through collaborative intelligence. 
However, the topology of these systems--how agents in MASs should be configured, connected, and coordinated--
remains largely unexplored. 
In this position paper, we call for a paradigm shift toward \emph{topology-aware MASs} that explicitly model and dynamically optimize the structure of inter-agent interactions.
We identify three fundamental components--\textbf{agents}, \textbf{communication links}, and \textbf{overall topology}--that collectively determine the system's \textbf{adaptability}, \textbf{efficiency}, \textbf{robustness}, and \textbf{fairness}. 
To operationalize this vision, we introduce a systematic three-stage framework: 1) agent selection, 2) structure profiling, and 3) topology synthesis. 
This framework not only provides a principled foundation for designing MASs but also opens new research frontiers across language modeling, reinforcement learning, graph learning, and generative modeling to ultimately unleash their full potential in complex real-world applications. 
We conclude by outlining key challenges and opportunities in MASs evaluation. 
We hope our framework and perspectives offer critical new insights in the era of agentic AI.
\end{abstract}


\section{Introduction}
Large Language Model (LLM)  agents~\citep{wu2023autogen,li2023camel,liu2025advances} have demonstrated strong capabilities across diverse domains such as code generation~\citep{hong2024metagpt,islam2024mapcoder,qian2023chatdev}, mathematical reasoning~\citep{gou2023tora,xie2024mathlearner}, and social simulation~\citep{huang2024adasociety,zhao2024competeai}. Building on the impressive capabilities of single-agents in general tasks, \textbf{LLM-based Multi-Agent Systems (MASs)}~\citep{ijcai2024p890} have emerged as a powerful paradigm that coordinate multiple specialized agents--e.g. planners, retrievers, reasoners--to collaboratively solve complex, composite tasks. 
These systems assign distinct roles, such as planners, retrievers, and reasoners, to distinct agents, enabling human-like collaboration through a division of labor. 
Similar to a human organization, 
a MAS operates as a \emph{dynamically evolving ensemble} that continually reconfigure to maintain a \emph{moving equilibrium}. As agents join, depart, or evolve, the system adapts its collaboration structure to optimize for long-term performance under resource and latency constraints. 

To enable inter-agent communication, coordination, and role specialization, MASs must support agents that are not only individually competent but also capable of collaborating through well-orchestrated interaction patterns.
While individual agent competence has advanced rapidly, the organizational principle of \textbf{how agents should be configured, connected, and coordinated--its topology}--remains underexplored~
~\citep{qian2025scaling}. 
This reveals an urgent need to rethink MAS design: 
\emph{What topological structures of LLM-based MASs are best to accomplish an unseen, complex task?}

\paragraph{Why Topology Matters?}
In MASs, inter-agent communication determines how information flows, how tasks are decomposed, and how consensus are reached. 
As task complexity increases and the number of agents scale up, communication costs often rise quadratically~\cite{qian2025scaling}. 
Na\"ive scaling--adding more agents without adaptive structural design--often results in redundant communication, higher GPU memory and token costs, longer response latency, and suboptimal coordination~\citep{qian2025scaling,zhang2024cut,wang2025agentdropout}. 
\begin{figure}
    \centering
    \begin{subfigure}[t]{0.48\linewidth}
        \centering
        \includegraphics[width=\linewidth]{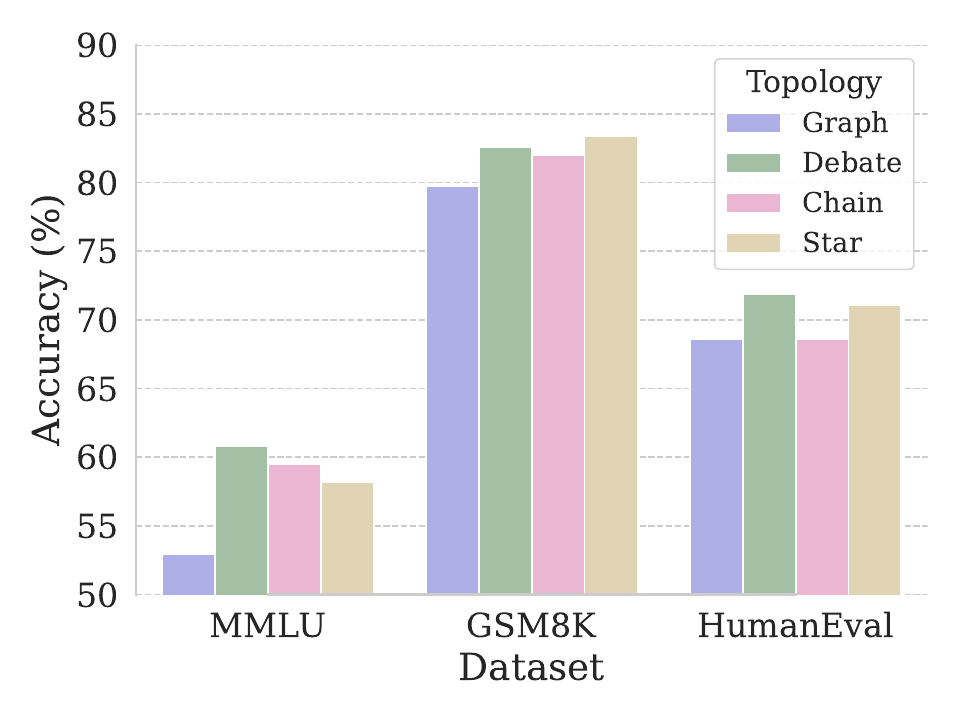}
        \vspace{-2em}
        \caption{GPT‑3.5}
        \label{fig:sub1}
    \end{subfigure}
    \begin{subfigure}[t]{0.48\linewidth}
        \centering
        \includegraphics[width=\linewidth]{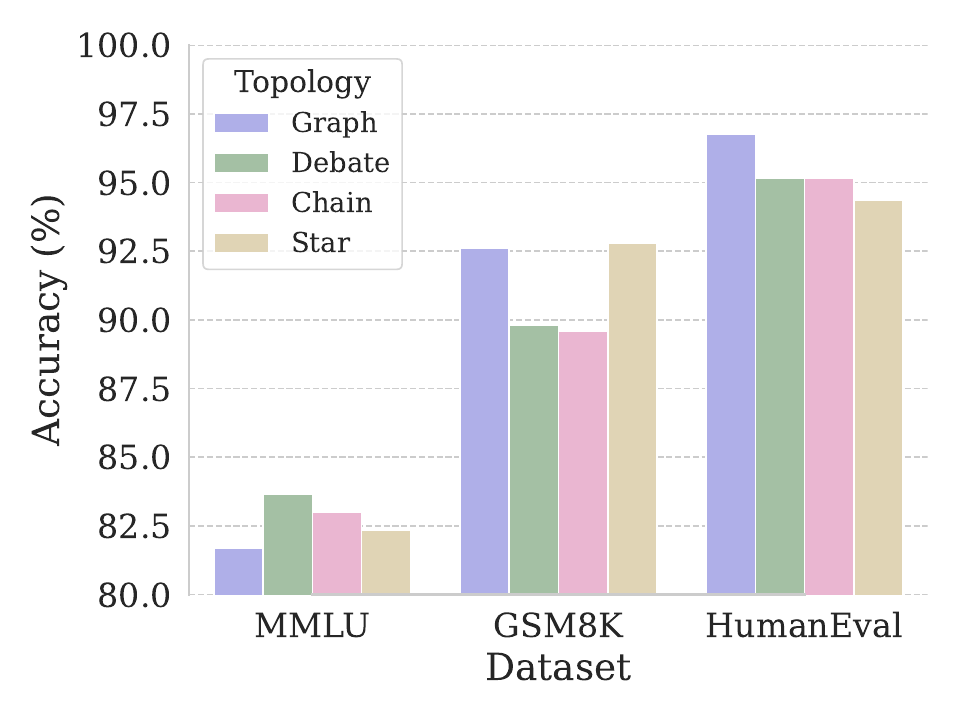}
        \vspace{-2em}
        \caption{GPT‑4o}
        \label{fig:sub2}
    \end{subfigure}
    \vspace{-1em}
    \caption{Comparison of topological structures across datasets. Results show that various tasks may require a specific topological structure.}
    \label{fig:parallel}
    \vspace{-5mm}
\end{figure}

Beyond \textbf{efficiency} concerns, topology also directly impacts \textbf{task performance}. As illustrated in \Cref{fig:parallel}, performance can vary by up to 10\% across different topological structures. Furthermore, the choice of communication structure carries critical safety implications,
whereby poorly designed topology may lead to misinformation propagation, biased reasoning, or the generation of harmful content by agents~\cite{yu2024netsafe}.


\paragraph{Challenges}
First, \emph{the optimal topology is task-dependent.} 
Universal topology design is difficult as structural preferences vary across domains--chain topologies are most effective for sequential workflows such as software development~\cite{zhang2024chain,hong2023metagpt}, while simulation or reasoning tasks favor tree or star topology that support information aggregation~\cite{qian2025scaling,wu2023autogen}. 
These dependencies are implicit, context-sensitive, and hard to generalize across tasks. 
Although prior works explore dynamic topologies~\citep{yue2025masrouter,liu2024dynamic}, they remain preliminary and still lack the ability to adaptively reconfigure topological structures in response to task goals or query contexts. 
Second, \emph{the design space of topological structures is inherently combinatorial.} agent selection, communication edges, and role dependencies grow exponentially with scale, rendering exhaustive optimization infeasible. 
This makes efficient structure optimization computationally prohibitive as the system scales.
Structure pruning techniques like edge pruning~\citep{zhang2024cut} and agent pruning~\citep{wang2025agentdropout,liu2024dynamic} are often applied in a static, one-off manner, lacking the capacity to model the dynamic interdependencies and evolving interactions among agents. 
Third, \emph{evaluating topologies requires end-to-end execution under real task dynamics}, which is computationally expensive and requires in-depth analytic modeling. Unlike in single-agent settings, the effectiveness of a given topology cannot be assessed in isolation. Its performance must be measured through full system execution within specific task contexts. 

\paragraph{What Should be Learned and How?}
We summarize three core components in MASs: \textbf{agents}, \textbf{communication links}, and the \textbf{overall topology} that governs how these agents interact. 
First, the \emph{candidate agent set} should consist of capable and complementary specialists tailored to the task--balancing diversity of expertise with coordination efficiency. 
Second, \emph{communication links} should be optimized to eliminate redundant messages and minimize communication overhead. 
Third, the \emph{topological structure} must align with the task's coordination pattern--whether hierarchical, modular, or distributed--to enable smooth information flow and balanced participation among agents. 

\begin{figure}
\centering{\includegraphics[width=0.98\columnwidth]{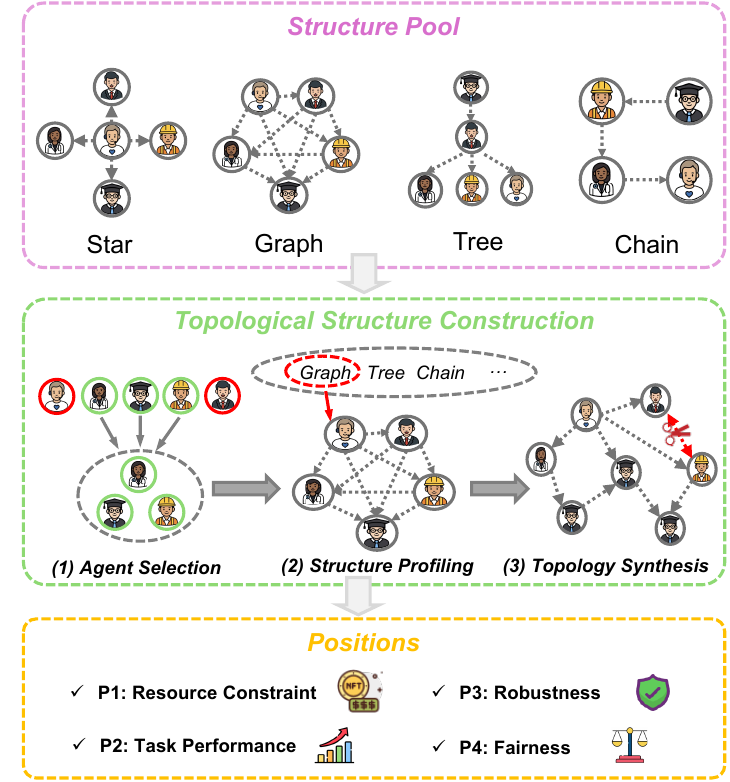}}
\vspace{-0.1in}
\caption{The proposed three-stage framework: agent selection, structure profiling, and topology synthesis, which grounds four key positions.}
\vspace{-0.2in}
\label{fig:roadmap}
\end{figure}

\paragraph{Positions} 
In this paper, we advocate a \emph{structure-aware paradigm that treats topological design as a top priority LLM-based MASs}. We propose a unified framework (\Cref{fig:roadmap}) that decomposes topological structure optimization into three stages: agent selection (Section~\ref{sec:agent_selection}), structure profiling (Section~\ref{sec:structure_profiling}), and topology synthesis (Section~\ref{sec:topology_systhesis}). 
This perspective reframes topology as a learnable, dynamic property--rather than a static configuration--opening new directions toward more \textbf{adaptive, efficient, robust, and fair} multi-agent systems.
\paragraph{Broader Impact}
Current agent-based LLM applications can potentially benefit from a topology-aware structure.
For instance, in customer support automation, dynamic agent topologies can reduce response latency, minimize redundant information routing, and improve relevance in multi-turn dialogues for support systems involving billing, troubleshooting, and multilingual hand-offs.
For collaborative code generation, task-specific topologies can support better coordination among agents with distinct roles (\eg planner, coder, tester), which adapt as tasks evolve.
Moreover, in scientific discovery, research assistants powered by LLMs can benefit from topologies that optimize how domain-specific agents (\eg literature retrievers, experimental planners, hypothesis evaluators) interact, particularly when solving open-ended scientific queries.
Given the dynamic and flexible nature of interaction topology, more applications will emerge and benefit.

The organization of this paper is as follows: we first introduce the basics of LLM-based MASs and give the position statements in \Cref{sec:position}.
We then attempt to argue positions we stated in the three-step framework of topological structure learningin \Cref{sec-framework}. 
Subsequently, \Cref{sec:challenges} discusses the key challenges that explain why current MASs still fall short of realizing these topology-aware principles in practice.
Finally, \Cref{sec-conclusion} concludes the paper.
Note that due to space limitations, more details of our proposed research directions are in the appendix.


\section{Positions}\label{sec:position}

\subsection{Definition of LLM-based Multi-Agent Systems}
LLM-based single-agent systems excel on a wide range of tasks, but can struggle with long-horizon, tool-intensive, or multi-step problems. 
This motivates the development of MASs, where multiple LLM-based agents collaborate through natural language communication to constructively divide roles, specialize functions, and jointly solve tasks.
To support such collaboration, we define an LLM-based MAS as a set of autonomous agents instantiated from LLMs, denoted as $\hat{\mathcal{A}} = \{a_1, ..., a_i\} \subseteq \mathcal{A}$. Each agent $a_i \in \mathcal{A}$ is assigned a specific role, such as planner, coder, or verifier, and interacts with other agents to collectively solve tasks.
\Cref{sec-append-notation} shows our notations. 

In practice, MASs adopt various structural topologies, such as chains, stars, trees, or densely connected graphs, each encoding different reasoning workflows and communication costs.
For instance, chain structures enable step-by-step refinement (\eg MetaGPT~\citep{hong2023metagpt}), while tree structures support hierarchical planning (\eg AutoGen~\cite{wu2023autogen}). Star topologies simplify control via centralized agents (\eg CAMEL~\citep{li2023camel}), and fully connected graphs maximize information sharing (\eg GPTSwarm~\citep{zhuge2024gptswarm}).
Each structure presents distinct trade-offs in coordination efficiency, scalability, and reasoning capability, and is inherently better suited to different tasks depending on its specific communication patterns and collaborative requirements.
Formally, we model the collaboration structure as a directed graph $\mathcal{G} = (\mathcal{A}, \mathcal{E})$, where $\mathcal{E} \subseteq \mathcal{A} \times \mathcal{A}$ represents the communication links.
At each round $t$, agent $a_i$ receives messages from its in-neighbors and generates an updated response
$R_i^{(t)} = a_i(P_{\text{sys}}, P_{\text{user}}^{(t)})$, where $P_{\textit{sys}}$, $P_{\text{user}}$ denote the system prompt and user prompt at round $t$ respectively.
$P_{\text{user}}^{(t)} = P_{\text{user}}^{(0)} \cup \mathcal{O}_i^{(t-1)}$,
where $\mathcal{O}_i^{(t-1)}$ denotes the aggregated messages from neighbors of agent $a_i$.

\subsection{Position Statement}

\begin{position}
\textbf{(Resource Constraint)}
The topological structure in MASs directly determines the resource required for agent collaboration
\end{position}
Poorly designed topologies such as fully connected graphs lead to redundant communication, increasing VRAM usage, and increased latency, making collaboration costly, whereas centralized topologies such as star structures concentrate workloads on a minority of nodes, leading to computational bottlenecks.

\begin{position}
    \textbf{(Task performance)}
    The topology of agents interactions serves as a critical factor in MASs that can either amplify or hinder task performance.
\end{position}
Different applications favor different agent structures for better performance.
For instance, collaborative coding (\eg ChatDev~\cite{qian2023chatdev} and AutoCodeRover~\cite{zhang2024autocoderover}) often uses chain structures for task decomposition, while world simulation (\eg Minecraft~\cite{suwannik2022minecraft} and ALFWorld~\cite{shridhar2021alfworld}) adopts a tree-based structure for better coordination.

\begin{position}
    (\textbf{Robustness})
    The robustness of MASs critically depends on their topological structure, which shapes the system’s tolerance to failures and adversarial perturbations.
\end{position}
Topologies with redundancy or modular design can sustain failures and resist misinformation. For example, debate structures introduce multiple verification paths that mitigate single-point failure.
In contrast, fragile topologies amplify risks of collapse and error propagation, such as centralized star structures.

\begin{position}
    \textbf{(Fairness)}
    The topology of agent interactions in MASs significantly influences the fairness of resource allocation and decision-making processes.
\end{position}
Balanced topologies enable diverse roles and personas (\eg planners, critics, verifiers) to participate equitably, while centralized structures risk privileging certain agents and marginalizing others, leading to unfairness issues.

\section{Position Arguments}
\label{sec-framework}
\vspace{-0.1in}

To verify the four positions (\textbf{P1–P4}) we advocated, we ground our arguments in the three-step framework of topological structure learning: \textit{agent selection}, \textit{structure profiling}, and \textit{topology synthesis}. 
At each stage, we discuss how topological structure considerations shape economic burden, task performance, robustness, and fairness. 
Beyond articulating these positions, we further outline potential research directions, demonstrating how each position can translate into actionable agendas for advancing MAS topology design.

\subsection{Three-Stage Framework Overview}
To better substantiate the stated positions, we introduce a unified framework for learning the optimal topological structure of MASs and argue each position through the lens of its corresponding stage.
The framework consists of three sequential stages: agent selection, structure profiling, and topology synthesis, as illustrated in Figure~\ref{fig:roadmap}.
Each stage has the potential to trigger new open problems and research directions.
Specifically, \ding{182} Given a new task, a pool of candidate agents with diverse capabilities is instantiated and made available. The first step is to select a subset of agents $\hat{\mathcal{A}} \subseteq \mathcal{A}$ best suited to collaborate on the task, based on factors such as skill specialization, role diversity, and past performance. This determines ``who'' participates in the communication topology.
\ding{183} For each task, the system identifies a \textit{macro structure} as a communication pattern, which captures abstract interaction forms, such as chains, stars, or trees. These macro structures reflect coarse-grained collaboration strategies and serve as an initial prototype for the following topology synthesis.
\ding{184} Based on steps \ding{182} and \ding{183}, the final stage synthesizes a concrete and directed communication graph, which we reference as the \textit{micro structure}.
It specifies the exact flow of information among the selected agents under the given macro structure, and determines which agents exchange messages and in what direction.

\vspace{-1mm}
\subsection{Agent Selection}
\label{sec:agent_selection}
\vspace{-1mm}

Modern AI platforms (\eg \href{https://www.together.ai/}{TogetherAI} 
and \href{https://huggingface.co/}{HuggingFace}) offer thousands of LLMs that can participate in MAS, varying in modality, capability, latency, parameter size, and cost.
We summarize two major problems in selecting agents to ensemble an effective and efficient system.

\begin{openproblem}
    Selecting proper agents is a cost-utility optimization problem that must balance \emph{exploration} (testing alternatives) and \emph{exploitation} (using the best‑known option) (P1).
\end{openproblem}

Users must decide \textit{which} agents to choose as candidates--high-performance but expensive SaaS LLMs versus smaller local checkpoints--and \textit{how} to allocate queries among agent.
Under-exploration risks missing models with superior task performance (P2), while over-exploration and exhaustive trials induce redundancy and costs (P1).

\begin{openproblem}
    Existing agent selection schemes are fundamentally limited by three major challenges: Single-Agent Focus 
    , Structural Myopia, and Scalability.
\end{openproblem}

Users must balance \textit{when} generalists and specialists fit into a particular task pipeline. 
We advocate a \emph{structure-aware}, \emph{lifecycle-aware} approach to agent selection that jointly optimizes multi-objective reward--task success, cost, latency, safety--while remaining adaptive to evolving agent availabilities and task distributions (P1, P2).

We propose two viable solutions, \textit{Static Selection} and \textit{Dynamic Selection}, depending on the specific scenarios. 
When task distribution, communication structure, evaluation metrics, costs, and the LLM catalog remain unchanged over the deployment horizon, \eg\ nightly batch summarization with a locked‑in set of SaaS APIs, where the utility of an agent can be seen as stationary, the problem reduces to estimating and selecting an optimal subset based on the interaction structure, and thus \textit{Static Selection} is preferred.
Formally, assuming that each agent in the pool is associated with a cost $C(a_i) \in \mathbb{R}_{\geq 0}$ (\eg API credits, GPU runtime), for a downstream task $x$ under macro structure $M$, we can define a utility function $\mathcal{U}:(x,M,\hat{\mathcal{A}}) \longmapsto [0,1] 
$ to evaluate the collective performance of a subset $\hat{\mathcal{A}} \subseteq \mathcal{A}$ on task $x$.
This utility can reflect metrics such as answer accuracy, semantic relevance, or user engagement (\eg click-through rate). 
The goal of static selection to find $k$ agents to balance performance-cost trade-offs: 

\vspace{-3mm}
\begin{equation}
\begin{aligned}
\mathcal{A}^\star
=
\underset{\hat{\mathcal{A}}\subseteq\mathcal{A},
|\hat{\mathcal{A}}|=k}{\mathrm{argmax}}
& \Bigl[
\mathbb{E}_{x\sim\mathcal{D}}
\bigl[\mathcal{U}(x,M,\hat{\mathcal{A}})\bigr] \nonumber \\
&\quad -
\sum_{a_i\in\hat{\mathcal{A}}}
\gamma(a_i) C(a_i)
\Bigr],
\end{aligned}
\label{eq:static-objective}
\vspace{-2mm}
\end{equation}
where $\gamma_i \in \mathbb{R}_{\geq 0}$ is a per-agent cost multiplier that emphasizes economy ($\gamma_i\uparrow$) or accuracy ($\gamma_i\downarrow$).

\begin{researchdirection} 
Incorporate topology-aware perspectives to frame agent selection as a structured reasoning problem, enabling efficient routing, selective activation, and dynamic composition.
\end{researchdirection}

\begin{researchdirection}
Integrate topology-aware perspectives to transform selection into a structured reasoning problem. which enables more efficient routing, selective activation, and dynamic composition.
\end{researchdirection}

\begin{researchdirection}
The multiplier $\gamma$ in Equation~\ref{eq:static-objective} can be treated as a stochastic, context-dependent signal, enabling adaptive selection policies that dynamically balance performance and economy as conditions drift. 
\end{researchdirection}

Multi‑armed bandits (MAB) and their combinatorial variants offer a principled framework to efficiently explore promising agents and exploit high-utility subsets. 
Unlike reinforcement‑learning (RL) training loops that continuously update a policy's parameters, MAB treats each agent as an \emph{immutable} stochastic arm, where one can query the arm (\ie\ a model), observe a reward, and pay its cost, but cannot directly alter its internal parameters. 
We provide a specific implementation in Appendix~\ref{app:static}.
Instead, in evolving environments, agent capabilities, task requirements, and agent interactions (\eg critique-revise chains) change over time, which favors \textbf{Dynamic Selection}.  
The \emph{marginal value} of an agent is contingent on 1) its own characteristics (capability, latency, cost), 2) the temporal context (current quotas, tool availability), and 3) which other agents are currently active.
%
%
%
Specifically, the overarching objective for dynamic selection is to learn a set function
$\mathcal{U}(\hat{\mathcal{A}},x)\in\mathbb{R}$ that estimates the \textbf{utility} of invoking a subset of agents $\hat{\mathcal{A}}\!\subseteq\!\mathcal{A}$ on an incoming task~$x$, given the total budget $B$: 
\vspace{-2mm}
\begin{equation}
\max_{\hat{\mathcal{A}}\subseteq\mathcal{A}} \mathcal{U}(\hat{\mathcal{A}},x),\quad\text\ C(\hat{\mathcal{A}})\le B.
\end{equation}

\begin{researchdirection}
Develop comprehensive task banks that provide fine-grained coverage of task types, difficulty levels, and required agent capabilities, enabling the learning of rich, transferable graph representations for effective agent selection.
\end{researchdirection}

\begin{researchdirection}
    Developing principled initialization strategies for node embeddings to enable effective graph representation learning and designing readout mechanisms to select agent subsets based on the learned embeddings.
\end{researchdirection}

In Appendix~\ref{app:dynamic}, we propose several options for the construction of task banks and fundamental features for the initialization of the embedding. This dynamic and inter-dependent setting can be naturally modeled by temporal graph neural networks (TGNNs)~\citep{rossi2020temporal, kumar2019predicting}. 
Each agent is a node, and each time-varying edge denotes an interaction (\eg routing a sub‑query or forwarding a partial answer) that arrives with a timestamp. 
TGNNs maintain a latent state for every node that is updated only when events involving that node occur, making them ideal for sparse, asynchronous communication patterns that dominate large-scale MAS.

We model agent interactions as a temporal graph $\mathcal{G}$,
represented by a sequence of time-indexed graph snapshots $\mathcal{G} = \{\mathcal{G}_t \mid t=1, \ldots,\mathcal{T} \}$. 
At training time, a task $x$ is sampled from $\mathcal{X}$. 
The agent selection algorithm in response selects a candidate subset $\hat{\mathcal{A}}$ (\ie\ agents that) for $x$ and constructs $\mathcal{G}$. 
At each time $t$, an active snapshot $\mathcal{G}_t = (\mathcal{V}_t, \mathcal{E}_t)$ 
specifies the \emph{participating} agents (\ie nodes) 
$\mathcal{V}_t = \{v_1, v_2, \dots, v_{n^t}\} = \hat{\mathcal{A}}^{t} \subseteq \hat{\mathcal{A}}$, along with their communication pathways $\mathcal{E}_t$,
where an edge $e_{ij}^{t} \in \mathcal{E}_t$ indicates 
an interaction between agent $i$ and $j$. 
Each edge induces a message $\mathbf{m}_{ij}^{t}$, which is aggregated by node $v_i$ from its neighborhood $\mathcal{N}_i$ to update its embedding $\mathbf{v}_i^t$:
\begin{align}
\mathbf{m}_{ij}^{t} &=\operatorname{Message}\left(\mathbf{v}_i^{t-1}, \mathbf{v}_j^{t-1}, \Delta t, \mathbf{e}_{i j}^{t}\right), \\
\mathbf{z}_i^{t} &= \mathrm{Agg}\left( \{ \mathbf{m}_{i j}^t, j \in \mathcal{N}_i \}  \right), \\ 
\mathbf{v}_i^{t} &= \mathrm{Update} (\mathbf{z}_i^{t}, \mathbf{v}_i^{t-1}),
\end{align}
where any node embedding $\mathbf{v}_i^t$  is updated only if agent $i$ participates in an interaction (edge) event, ensuring computational efficiency under sparse and asynchronous communication patterns. 

At the final timestamp $\mathcal{T}$, a permutation-invariant read-out module maps the node embeddings ${\mathbf{v}_i^{\mathcal{T}}}$ of $\hat{\mathcal{A}}$ and the task embedding $\mathbf{x}$ to a utility estimate: 
\vspace{-3mm}
\begin{equation}
\mathcal{U}(\hat{\mathcal{A}},x)=\mathrm{Readout}\left(\sum_{a_i\in\hat{\mathcal{A}}} \left[\mathbf{v}_i^{\mathcal{T}} \Vert \mathbf{x}\right]\right),
\end{equation}
where $\Vert$ is concatenation. 
The subset $\hat{\mathcal{A}}$ is scored via automatic or LLM-based evaluation as a function of its utility.  
Both the TGNN and the read-out module are trained jointly using a pairwise ranking loss that promotes higher scores for subsets yielding greater utility. The model learns to predict both the subset utility and each node’s marginal contribution. 

A fairness-aware read-out module can be introduced to prevent systematic bias toward resource-rich or proprietary agents. Without such correction, the model might consistently prefer high-capacity agents (e.g., GPT-4-class LLMs) that dominate short-term utility estimates, while overlooking smaller open-source agents that provide complementary reasoning diversity or cost efficiency (P4). 

Along this direction, the readout can incorporate group-level regularization or exposure balancing---e.g., ensuring that subsets include a mix of agent types (proprietary vs. open-source), or applying a mild penalty when all selected agents belong to a single high-cost group.

\vspace{-0.3em}
\subsection{Structure Profiling}
\label{sec:structure_profiling}
Empirically, different tasks exhibit distinct coordination patterns. For example, sequential workflows align with chains, hierarchical exploration with trees, centralized tool routing with stars, and high-bandwidth ideation with denser graphs~\cite{zhang2024chain, yao2023tree, besta2024graph}. The \textit{macro-level} topological structure in MASs directly determines the economic burden of collaboration. Imagining all agents in MASs adopt a fully connected structure, each round of interaction generates a large volume of redundant message exchanges, which not only increases token consumption but also leads to significant latency~\cite{wu2023autogen}.
This issue becomes even more pronounced as the system scales, where MAS faces the risk of exponential growth in token usage. For example, experiments demonstrate that unoptimized multi-round fully connected MAS can consume tens of times more tokens than single-agent or chain-structured systems~\cite{wang2025agentdropout}. 
Committing early to a macro-level archetype gives the system a strong inductive bias before any micro-level wiring, improving efficiency, robustness, and interpretability across tasks. We frame this explicitly as selecting among a discrete set of topological archetypes conditioned on the task, rather than directly optimizing the full edge set. Archetype selection injects the right inductive bias upfront (e.g., depth for decomposition, breadth for exploration), which stabilizes subsequent micro-search by bounding the feasible graph family and reducing non-stationarity in edge learning. Therefore, given a task $x \in T$, the goal is to select a single macro topology $M$ from a pre-defined structure pool  $\mathcal{M}$ (e.g., chain, tree, star, ring, fully-connected) that best trades off expected task reward and communication budget~\cite{wang2025agentdropout, besta2024graph}. We adopt a cost-aware criterion to pick the macro topology:
\begin{equation*}
    \vspace{-0.5em}
    M^{*} \;=\; \arg\max_{M \in \mathcal{M}} \;\; \mathcal{R}(x,M) \;-\; \lambda \, C(M),
    \vspace{-0.5em}
\end{equation*}
where $\mathcal{R}(x,M)$ estimates expected task reward under topology $M$ and $C(M)$ aggregates communication-related costs (e.g., token, latency), with $\lambda$ balancing quality and economy. The pool should be a small, human-interpretable set and can be expanded with DAG variants or layered stars when justified by use cases~\cite{zhang2024chain}.

For structure profiling, a compact classifier $f_{\theta}(\phi(T)) \to \mathcal{M}$ on (task, structure) pairs curated from prior runs, simulations, or modest expert annotation should be a feasible way, where task signature $\phi(T)$ is a descriptor derived from the prompt, tools, constraints, or historical traces (e.g., need for depth or breadth, central routing or peer debate). Using cross-entropy with cost-aware reweighting to reflect $C(M)$ (e.g., penalizing dense graphs) should also be considered~\cite{wu2023autogen, chen2023agentverse}. The output will be $M^*$ with possible macro hyper-parameters, such as max degree, allowed diameter, or layer count).
\begin{openproblem}
    How do we learn mappings from task signatures to a small set $\mathcal{M}$ that transfer across domains, tools, and models? We may need strategies that leverage weak supervision, priors, and continual updates without drifting. \textbf{(P1, P2)}
\end{openproblem}
\begin{openproblem}
    Different macro-level topological structures in MASs can lead to drastically different communication costs. How can we control the resource cost while maintaining the task performance? \textbf{(P1)} 
\end{openproblem}
\begin{researchdirection}
    Design few-shot structure classifiers with uncertainty. The compact classifiers can be trained on modest (task, archetype) corpora, with uncertainty gating to trigger budgeted probes only when needed; continual learning from deployment logs.
\end{researchdirection}

The robustness of MASs critically depends on their topological structure, which shapes the system’s tolerance to failures and adversarial influences~\cite{wu2023autogen, yao2023tree}. For example, tree structures may provide cross-verification, allowing the system to recover when individual agents fail or inject misinformation. In contrast, chain structures are fragile, as the failure of a single agent can break the reasoning flow, and adversarial perturbations introduced early can propagate unchecked through subsequent steps~\cite{du2023improving, hong2023metagpt}. These observations underscore that robustness is inherently a \textit{macro-level} topological property, aligning with our position \textbf{P3}.

\begin{openproblem}
    What redundancy is minimally sufficient to resist agent failures and adversaries without prohibitive overhead, and how do macro choices affect fair access to central nodes (e.g., star) and bias amplification?~\textbf{(P3, P4)}
\end{openproblem}
\begin{researchdirection}
    Benchmarks will be valuable and should annotate macro labels as well as report metrics by archetype class so we can attribute gains to topology rather than model scale or prompt tricks.
\end{researchdirection}
The structure profiling stabilizes subsequent micro-search, and enables principled cost–quality trade studies. Next, Section~\ref{sec:topology_systhesis} will instantiate edges and communication policies strictly within those bounds.

\subsection{Topological Structure Synthesis}
\label{sec:topology_systhesis}

While agent selection and structure profiling identify the macro-level pattern of interactions, synthesizing a micro-level communication topology is a pressing problem to further enhance the communication efficiency and task performance. 
Specifically, agent selection determines \textit{who} participates in the collaboration, and structure profiling identifies the \textit{macro-level pattern} of interaction. Effectively executing the task requires synthesizing a fine-grained communication topology that governs the actual flow of information between agents.
This \textit{micro-level} structure plays a critical role in determining not only which agents should communicate but also the edge directions, weights, and potential hierarchy (\eg tree depth or breadth). These decisions directly influence how efficiently information propagates throughout the system, ultimately shaping task performance~\citep{wang2025agentdropout}, communication cost~\citep{wang2025agentdropout,zhang2024cut}, and the robustness of the multi-agent collaboration~\citep{yu2024netsafe}.

\begin{openproblem}
    Fully connecting every pair of agents leads to redundancy and high communication costs: How do we prune unnecessary links while preserving task performance? \textbf{(P1)}
\end{openproblem}
Recent efforts have explored topological structure optimization in MASs through topology pruning-based methods, which aim to reduce redundancy and improve efficiency by selectively removing communication links from the topological structure. AgentPrune~\citep{zhang2024cut} learns a sparse spatial-temporal communication graph by optimizing differentiable edge masks with policy gradients and performs one-shot pruning to remove redundant message-passing edges while preserving overall task performance. 
Similarly, G-Designer~\citep{zhang2024g} adopts a reinforcement learning-based framework to learn task-specific sparse communication graphs by selecting critical communication links. AgentDropout~\citep{wang2025agentdropout}, in contrast, adaptively drops low-utility intra- and inter-round edges to compress communication graphs and reduce token usage during inference. GPTSwarm~\citep{zhuge2024gptswarm} regards agents as nodes in a computational graph and optimizes inter-agent communication links, enabling end-to-end learning of sparse collaboration topologies tailored to task requirements.
However, most of them suffer from limited scalability and high computational cost, leaving an open problem that urgently needs to be addressed in future research, which directly verifies our position \textbf{P1}.

\begin{openproblem}
    Different agent roles should occupy specific positions within the topological structure: How do we design the micro-structure while considering the specific role of each agent to maximize performance, make it more robust, while maintaining the fairness?~\textbf{(P2, P3, P4)}
\end{openproblem}
While many existing approaches focus on the communication connectivity among agents~\cite{zhang2024cut}, they often ignore the problem of the different, specialized roles of agents that should occupy distinct and meaningful positions within the topology. For example, a manager should be in the central position for overall coordination, while a verifier may serve best at the end of the pipeline. Therefore, this open problem implies the position statement of \textbf{P2}. 

In addition, several existing studies have begun to explore the robustness of different topological structures in MASs~\cite{yu2024netsafe}. For example, in some cases, a manager placed at the center can maintain coordination even if peripheral agents fail, while a verifier positioned at the end of the pipeline ensures that adversarial perturbations do not propagate unchecked.
Thereby, once the specific role of each agent is aligned with the task’s requirements, robustness emerges as a central challenge, which is exactly the concern articulated in our position \textbf{P3}.

Lastly, fairness is also deeply intertwined with the placement of agent roles. If certain roles (\eg managers or planners) are always granted central or high-degree positions, while others (\eg verifiers or critics) are pushed to the periphery, the system risks privileging dominant voices and marginalizing supporting ones. A fair topology should instead ensure that diverse roles and personas have equitable opportunities to contribute, preventing structural biases from distorting decision-making. This directly connects to our position \textbf{P4}.

\begin{researchdirection}
    Measuring the fine-grained, task-specific influence of individual communication links is essential for optimizing \textit{micro-level} topological structure of MASs.
\end{researchdirection}
\vspace{-0.5em}
A central challenge in optimizing \textit{micro-level} topologies in MASs is to assess each communication link’s contribution to task performance. Beyond coarse assumptions (\eg fully connected), fine-grained evaluation enables pruning strategies that reduce overhead while preserving accuracy. Counterfactual reasoning estimates link utility from performance drops upon removal, while scalable approximations such as sampling or gradient sensitivity offer efficiency. Since links interact, group-level perturbations capture collective contributions that individual assessments may miss.
\Cref{appendix:topology_optimization} provides further details on potential implementations of this research direction.

\begin{researchdirection}
    Exploring generative models to synthesize task-specific communication topologies offers a flexible pathway to automate and adapt multi-agent coordination structures.
\end{researchdirection}
Pre-trained language models can directly generate \textit{micro-level} topologies from task descriptions, avoiding costly link-level evaluation. Likewise, graph generative models conditioned on agent roles can produce role-aware interaction structures that capture specialization and interdependence.

\section{Challenges and Discussions} \label{sec:challenges}
\vspace{-0.5em}
The positions and research directions articulated above are both compelling and necessary. Yet, despite their clarity, most MASs still fall short of realizing them in practice.
We believe the following constitute the principal barriers:

\noindent \textbf{\textit{Developing Topology-Aware Benchmarks.}}
Current benchmarks focus on general reasoning or coding tasks and fail to capture topology-related dynamics such as agent scaling, communication redundancy, and error propagation.
Future work should design topology-aware benchmarks and metrics that reflect coordination efficiency and structure sensitivity.

\noindent\textbf{\textit{Real-World Robustness.}}
While many MASs achieve high benchmark scores, they often rely on static or contaminated data. Evaluating genuine reasoning capability under noisy, adversarial, or evolving environments remains difficult.

\noindent\textbf{\textit{Scaling and Reality Gap.}}
Most MASs are tested on small populations within simulators. Scaling to hundreds of agents or bridging the simulation-to-reality gap introduces latency, bandwidth, and stability issues that current setups overlook.

\noindent\textbf{\textit{Multi-System Coordination.}}
In real deployments, MASs span multiple organizations and data boundaries. This raises challenges in privacy-preserving topology learning, distributed optimization, and cross-system alignment.

\noindent\textbf{\textit{Trust and Accountability.}}
As MAS ecosystems grow, ensuring verifiable, trustworthy, and failure-resilient topologies becomes essential. Research on trust-aware pruning, structural verification, and explainable topology design is urgently needed.

\noindent\textbf{\textit{Expensive Optimization.}}
Most topology optimization methods for MASs are computationally expensive, such as relying on counterfactual approaches. Their cost grows rapidly with the number of agents and links, making it unrealistic.

\vspace{-0.3em}
\section{Conclusion}
\label{sec-conclusion}
\vspace{-0.3em}
This position paper advocates for topological structure learning as a critical and underexplored direction in LLM-based MASs. We have outlined a conceptual decomposition of the topology design problem into agent selection, structure profiling, and topology synthesis, each posing distinct algorithmic and theoretical challenges. To this end, we attempt to propose a systemic framework to solve these three sub-problem.
We hope this paper will catalyze future research into topological structure learning and lay the foundation for more intelligent and adaptive MASs.

\section{Limitations}
While this paper advocates for topological structure learning as a critical research priority for LLM-based MASs, our discussion remains primarily conceptual and forward-looking.
In addition, the proposed three-stage framework is summarized from patterns observed across most existing works, serving as a generalized abstraction rather than an exhaustive taxonomy. However, certain specialized or hybrid optimization methods may not fit neatly into this framework, which could limit the completeness of our argumentation for the stated positions.
Last but not least, our analysis primarily centers on LLM-based textual agents. Extending the framework to multimodal or embodied agents (\eg vision-language or robotics systems) may introduce additional complexities not addressed in this paper.

\section{Ethical Considerations}
Optimizing the topological structure of MASs may lead to new ethical risks beyond efficiency and performance.
Structural imbalances could inadvertently privilege certain agents, underscoring the importance of fairness-aware topology learning.
Besides, cross-system communication also requires careful treatment of privacy and governance boundaries.
\bibliography{references}

\newpage
\appendix
\section{Notations}
\label{sec-append-notation}

\Cref{tab-notation} shows the notations used thoughout the paper.

\begin{table}[htbp]
\centering
\caption{Notation used in this paper.}
\label{tab-notation}

\resizebox{.5\textwidth}{!}{
\begin{tabular}{ll}
\toprule
\textbf{Symbol} & \textbf{Meaning} \\
\midrule
$\mathcal{A} = \{ a_i \}$ & Set of candidate agents \\
$\mathcal{A}^* \subseteq \mathcal{A}$ & Optimal set of agents \\
$\hat{\mathcal{A}}$ & Selected agents, $\lvert\hat{\mathcal{A}}\rvert = k$ \\
$T = \{x\}$ & Specific task instance \\
$M$ & Macro‑structure / coordination pattern \\
$C(a_i)$ & Cost of agent $a_i$ \\
$\gamma$ & Cost–reward trade‑off coefficient ($\gamma \ge 0$) \\
$\mathcal{R}(x,M,\hat{\mathcal{A}})$ & Expected reward under $M$ with $\hat{\mathcal{A}}$ \\
$\mathcal{U}(x,M,\hat{\mathcal{A}})$ & $\mathcal{R} - \gamma \sum C(a_i)$ utility \\
$\mu(\hat{\mathcal{A}})$ & True expected utility of $\hat{\mathcal{A}}$ \\
$\mathcal{G} = (\mathcal{V}, \mathcal{E}, \mathcal{T})$ & Temporal graph \\
$\mathrm{Message}(\cdot)$ & Message function \\
$R_{i}$ & response from agents \\
$P_{sys}$, $P_{user}$ & system prompt, user prompt \\
$\mathcal{O}_i$ & aggregated messages from neighbor agents\\
\bottomrule
\end{tabular}
}
\vspace{-.2in}
\end{table}

\section{Algorithm}
Algorithm~\ref{alg:greedy_knapsack} presents the strategy of greedy selection under budget $B$ for agent selection.
\begin{algorithm}[H]
\caption{Greedy Selection Under Budget $B$.}
\label{alg:greedy_knapsack}
\begin{algorithmic}[1]
\Require Ground set $\mathcal{A}$, submodular $f(\cdot)$, cost $C(\cdot)$, budget $B$
\State $\hat{\mathcal{A}}\leftarrow\emptyset$, ; $b\leftarrow0$  \Comment{$b$ tracks accumulated cost}
\While{$\exists a_i \in \mathcal{A} \setminus\hat{\mathcal{A}}$ \textbf{s.t.} $b+C(a_i)\le B$}
\State $\hat{a} \leftarrow \underset{a_i\in\mathcal{A}\setminus\hat{\mathcal{A}}}{\mathrm{argmax}}
\frac{f(\hat{\mathcal{A}}\cup\{a_i\})-f(\hat{\mathcal{A}})}{C(a_i)}$
\If{$b + C(a_i) \le B$ \textbf{and} $f(\hat{\mathcal{A}}\cup\{a_i\})>f(\hat{\mathcal{A}})$}
\State $\hat{\mathcal{A}}\leftarrow\hat{\mathcal{A}}\cup\{a_i\}$; \quad $b\leftarrow b + C(a_i)$
\Else
\State \textbf{break} \Comment{No affordable beneficial addition remains}
\EndIf
\EndWhile
\State \Return $\hat{\mathcal{A}}$ \Comment{$(1-1/e)/2$ approximation~\citep{sviridenko2004note}}
\end{algorithmic}
\end{algorithm}


\section{Details of Agent Selection}

Following the formulation in Section~\ref{sec:agent_selection}, we can apply variations of MAB (Appendix~\ref{app:online_identification}), and design intermediary rewards to overcome the sparsity in reward signals (Appendix~\ref{app:intermediary_rewards}).

\subsection{Static Agent Selection} 
\label{app:static}
Once cost-constrained exploration stabilizes empirical rewards $\hat{\mu}_i$, the problem reduces to choosing a subset $\hat{\mathcal{A}} \subseteq \mathcal{A}$ that maximizes an ensemble utility $\mathcal{U}(\mathcal{A}, x)$ subject to the same budget.
An exhaustive search over all $\smash{2^{|\mathcal{A}|}}$ subsets is generally intractable \citep{chen2025optimizing}, but we can exploit the structure of $\mathcal{U}$. 
\emph{Diminishing-returns.} 
In ensembling~\citep{wang2024mixture}, 
the marginal benefit of adding one additional agent can diminish due to overlapping knowledge or capabilities. 
Consequently, $\mathcal{U}(x,M,\hat{\mathcal{A}})$ is approximately submodular in $\hat{\mathcal{A}}$--the benefit of adding an agent decreases as the committee grows. When this holds, the classic greedy algorithm—adding agents one at a time based on their marginal gain--achieves a $(1 - 1/e)$ approximation to the optimal selection. 

\emph{Budget-constrained Variant.} 
When the committee size is flexible but a hard budget $B$ exists, the classical ratio-greedy knapsack algorithm can be applied (Appendix Algorithm~\ref{alg:greedy_knapsack}). 
At each step, the algorithm selects the agent with the highest \emph{benefit-to-cost} ratio
$\bigl(\hat{\mu}_i - \gamma(a_i)C(a_i)\bigr)!/!C(a_i)$,
and continues until the total cost reaches the budget $B$. 
Finally, this approach naturally extends to \emph{hierarchical task decomposition}.  
When a complex task can be divided into smaller sub-tasks (e.g., planning, solving, verification), we run separate budgeted bandits and greedy selections for each sub-task. We then compose the selected agents into a full workflow, enabling fine-grained agent specialization while preserving global budget efficiency.

\subsection{Dynamic Agent Selection} 
\label{app:dynamic}
\noindent \textbf{Node Embedding Curation.} 
We propose constructing node embeddings $\mathbf{v}_i^0$ from four feature groups: 
1) \emph{Capability}: context length, training corpus size, parameter counts, embedding dimensions;
2) \emph{Cost}: price per 1000 tokens, hardware costs; 
3) \emph{Performance history}: success rate, latency, recent tasks metrics such as F1/BLEU scores~\citep{papineni2002bleu}; 
and 
4) \emph{Persona attributes}: communication styles, creativity, safety stance.

\noindent \textbf{Task Bank.} 
Tasks $\mathcal{X}$
are organized in a \emph{coarse‐to‐fine} faceted taxonomy:
1) \emph{skill}: reasoning, code generation, tool use;
2) \emph{modality}: text, code, image, audio, and video; 
3) \emph{interaction pattern}: single/multi-turn conversations or planning. 
Each task receives multiple labels across these axes, with associated seed prompts and evaluation rubrics. 

\noindent \textbf{Systematic Persona Synthesis.} 
Current approaches to agent selection often rely on handcrafted prompt templates to define personas~\citep{jin2024agentreview}. 
For example, \emph{creative} agents favor broader exploration, while \emph{conservative} agents prefer safe, high-confidence regions~\citep{sun2024llm}. 
Agents might prefer clarity (\emph{executive}), critical thinking (\emph{Socratic}), or story-telling (\emph{narrative}). 

However, LLMs can deviate from their assigned roles, even when provided with explicit instructions and behavioral criteria. 
This leads to behavioral homogeneity and communication redundancy across agents. 

To overcome this, future work could explore two complementary directions for persona synthesis: 
\begin{itemize}[leftmargin=1em]
    \item \emph{Parametric Persona Curation.} Rather than \emph{discrete}, natural-language-based templates, personas can be embedded in a continuous, interpretable latent space--parameterized along axes such as exploration-exploitation, verbosity, or risk tolerance. Trait-conditioned adapters or prompt generators can then enable flexible and fine-grained control by smooth interpolation between styles.
    \item \emph{Data-Driven Persona Discovery}. 
    Alternatively, we can mine diverse agent behaviors by prompting LLMs with varied instructions and clustering resulting dialogue patterns. Embedding these interactions reveals the range of emergent personas, which can then be selectively instantiated.
\end{itemize}

In both approaches, automated validation--through competency, bias, and efficiency checks--ensures that newly synthesized agents are both safe and effective before deployment~\citep{jin2024better}.



\textbf{Smooth Learning and Cold Start.} 
Deploying an under-initialized agent-selection algorithm in a new environment can be risky, incurring prohibitive cost, latency, or safety violations. 
Naive exploration can lead to excessive cost, latency, or even safety violations before effective learning. To ensure \emph{smooth learning}, one promising research direction is to design effective initialization strategies that provide low-regret starting points and enable safe, efficient adaptation from the outset.


\noindent\textbf{Online Identification via Cost-aware Bandits.}
\label{app:online_identification} 
Each candidate agent $a_i$ is modeled as a stochastic bandit arm with cost $C(a_i)$ and reward from a utility scoring function. 
Our objective is to identify a subset of agents that maximizes utility per unit cost across a stream of tasks. 
Variants based on existing algorithms, such as contextual bandits~\citep{chu2011contextual} or Neural UCB~\citep{zhou2020neural}, guide exploration using query features (\eg domain tag, length, etc.), 
prioritizing under-explored yet promising agents. Each round, the controller predicts the expected reward, 
applies an optimism bonus, and selects the top-$k$ arms. 
%
While feedback-graph bandits~\citep{alon2015online,cesa2021cooperative,rouyer2022near} considered structural interactions among arms,
they mostly focus on leveraging this topology to improve sample efficiency by assuming fixed and given graph structures rather than optimizing them.
Under a spend budget $B$, budget-aware variants based on \emph{budgeted-UCB}~\citep{ding2013multi} and \emph{budgeted Thompson Sampling}~\citep{xia2015thompson} 
manage agent selection by treating each pull as consuming cost $C(a_i)$ until budget $B$ is reached. 

\noindent\textbf{Intermediary Rewards via Automatic Scorers.} 
\label{app:intermediary_rewards} 
Human or gold-standard labels offer accurate utility assessments but are costly and sparse. 
Instead, automatic scorers can be used to provide instantaneous, low-cost reward estimates. 
Scorer models are architecturally agnostic and can take various forms: 
a \emph{smaller, instruction-tuned} specialist model~\citep{li2025mind,hu2024automated}, 
a rule-based evaluator (e.g. exact string match, BLEU/ROUGE, BERTScore, toxicity-detectors), 
or LLM-as-a-Judge~\citep{zheng2023judging}. 
Critically, the scorer does not need to surpass production agents in size or capability--it merely serves as a lightweight, cost-effective proxy for reward signals. 
As LLM-based automatic feedback might not always be accurate, we can develop superior mechanisms for providing corrective feedback or rewards to shape agent behaviors in real time without blocking the training flow. 

\noindent\textbf{Motivation for Temporal GNN for Agent Selection.}
\begin{itemize}[leftmargin=1em]
    \item \emph{Time-varying Connections}: Communication patterns between agents can change rapidly (e.g., requests for help, forwarding solutions, ephemeral collaborations). An agent may broadcast to many peers during peak load but fall back to a few trusted collaborators when quotas tighten.  TGNNs explicitly encode such time‐varying edges.
    \item \emph{Sequential Dependencies}: The impact of a message depends on \emph{when} it arrives: an expensive agent triggered late in the interaction window may violate a latency SLA even if its answer is accurate. TGNNs condition updates on inter‑event time $\Delta t$. 
    \item \emph{Path Dependencies}: Agents accumulate \emph{context} over rounds of communication. TGNNs enable long‐range modeling by preserving this information through their node memories.
\end{itemize}

\section{Details of Structure Profiling} 
\label{apx:span}
\subsection{Existing Studies}
Automated design of agentic systems is a nascent paradigm aiming to replace static, hand-crafted agent organizations with learned ones~\citep{hu2024automated}. AFlow~\citep{zhang2024aflow} reformulates an agent workflow as a programmatic graph and uses Monte Carlo tree search to iteratively refine this workflow. Similarly, MaAS~\citep{zhang2025multi} can sample a query-dependent agent configuration from this supernet at runtime.
MasRouter~\citep{yue2025masrouter} uses a cascaded controller network to dynamically decide how to route a user’s query through a bunch of agents and determine the collaboration mode.
G-Designer~\citep{zhang2024g} leverages a graph neural network to generate a communication topology for the agents based on the task at hand. Similarly, CommFormer~\citep{hu2024learning} uses a continuous relaxation of the adjacency matrix to allow gradient-based search of the optimal inter-agent communication links.
To guide the formation of an effective communication structure, AOP~\citep{li2024agent} emphasizes structural principles like solvability, non-redundancy and completeness, where meta agent decomposes and assigns agents. The reinforced meta-thinking agents framework ReMA~\citep{wan2025rema} has a high-level agent that plans and monitors the reasoning strategy, as well as a low-level agent that executes the reasoning steps. The existing work motivates developing a general method to determine optimal multi-agent structures adaptively.
\subsection{Future Discussion}
The framework above demonstrates that exploring adaptive structures that evolve during problem solving is a promising approach. That is, instead of fixing $\mathcal{G}$ from the start, agents could dynamically rewire their communication links based on intermediate results. 
Another direction is to incorporate heterogeneous objectives and game-theoretic considerations, e.g., in competitive or mixed cooperative-adversarial settings, the optimal structure might involve forming coalitions or introducing mediator agents. Our task can be extended to design structures that are robust against adversarial agents or that balance fairness and information sharing in a team. Furthermore, consider to integrate human insights and interpretability into structure profiling, i.e., the learned graphs and role assignments could be constrained to mirror effective human organizational strategies, such as hierarchical delegation and review committees, so that the resulting agent society is interpretable and trustworthy. 
Finally, as foundation models diversify, an exciting frontier is profiling structures across modality-diverse agents, e.g., LLMs alongside vision or robotics agents.
In summary, structure profiling for LLM-based agents is a rich field, and future work can build on this foundation by developing more adaptive, robust, and generalizable mechanisms for architecting multi-agent intelligence.

\section{Details of Topological Structure Synthesis}
\subsection{Details for Task-Aligned Utility Estimation for Topology Optimization}
\label{appendix:topology_optimization}
\noindent\textbf{Counterfactual Topology Intervention.}
To quantify the topological structure importance of each agent $a_i \in \hat{\mathcal{A}}$ or communication edge $e_{i} \in E$ within a synthesized topological structure $G = (\hat{\mathcal{A}}, E) \in \mathcal{G}_M(\hat{\mathcal{A}})$, we adopt a counterfactual reasoning approach. Specifically, we define the task-aligned contribution of each component, which can be either an agent $a_i$ or a communication edge $e$, as the drop in utility resulting from its removal. Let $\mathcal{U}(x, M, \hat{\mathcal{A}})$ denote the utility associated with executing task $x \in T$ under macro-structure M and agent subset $\hat{\mathcal{A}}$. 
Formally, we define the contribution of each agent $a_i \in \hat{\mathcal{A}}$ as
\begin{equation}
    \phi(a_i) := \mathcal{U}(x, M, \hat{\mathcal{A}}) - \mathcal{U}(x, M, \hat{\mathcal{A}} \setminus \{a_i\}),
\end{equation}
and the contribution of each edge $e_{i} \in E$ as
\begin{equation}
    \phi(e_{i}) := \mathcal{U}(G; x) - \mathcal{U}(G \setminus \{e\}; x).
\end{equation}
Here, $\mathcal{U}(G; x)$ denotes the utility under a given micro-topology G, and $G \setminus \{e\}$ or $\hat{\mathcal{A}} \setminus \{a_i\}$ respectively represent topologies with the edge or agent removed.
Given these contribution scores, we formalize the topology refinement as a constrained selection problem. The goal is to identify a subgraph $G' = (\hat{\mathcal{A}}{'}, E{'}) \subseteq G$ that retains only components with sufficiently high task-aligned utility, while maximizing the overall system performance.
Let $\delta_v$ and $\delta_e$ denote contribution thresholds for nodes and edges, respectively. Then, the resulting substructure is obtained by solving:
\begin{equation}
    \begin{aligned}
    G^* = \arg\max_{\hat{\mathcal{A}}{'}, E{'}} \, \mathcal{U}(G'; x),
    \: \text{subject to}  \\ \;\, \phi(a_i) \geq \delta_v, \forall \, a_i \in \hat{\mathcal{A}}{'};
    \;\, \phi(e) \geq \delta_e, \forall \, e_{i} \in E{'}.
    \end{aligned}
\end{equation}

\noindent\textbf{Efficient Estimation Approaches.}
To mitigate the computational burden of explicit counterfactual interventions, two efficiency-oriented approximations are explored that that estimate structural importance while preserving alignment with task utility.
The first, \textit{Marginal Contribution Sampling}, approximates the counterfactual utility drop by averaging over multiple stochastic subgraph configurations. Instead of removing one agent or edge at a time, it estimates the marginal gain of including a component by evaluating its incremental effect across sampled subsets $S_j \subseteq \hat{\mathcal{A}}$, as formalized by $\phi(a_i) \approx \frac{1}{m} \sum_{j=1}^m [\mathcal{U}(S_j \cup \{a_i\}) - \mathcal{U}(S_j)].$
In contrast, \textit{Gradient-Informed Attribution} leverages differentiability in the utility function to infer structural relevance via sensitivity analysis. By relaxing discrete structures into continuous soft masks (\eg edge weights $w_{e_i}$ or agent activation factors $\alpha_{a_i}$), it computes importance scores via backpropagation as $\phi(e_i) \approx \left| \frac{\partial \mathcal{U}}{\partial w_{e_i}} \right|$ or $\phi(a_i) \approx \left| \frac{\partial \mathcal{U}}{\partial \alpha_{a_i}} \right|$.

\noindent\textbf{Group-level Structural Effects.}
While the aforementioned approaches estimate structural importance at the level of individual agents or edges, they implicitly assume that each component contributes independently to task performance. However, the utility of an agent is tightly coupled with others in many cases, such as collaborative planning. To capture such interdependencies, we consider \textit{Group-level Perturbation}, which generalizes counterfactual intervention from individual units to agent clusters.
Concretely, we evaluate the collective utility drop incurred by removing a group of agents $\mathcal{S} \subseteq \hat{\mathcal{A}}$, defining their joint contribution as:
\begin{equation}
    \phi(\mathcal{S}) := \mathcal{U}(\hat{\mathcal{A}}) - \mathcal{U}(\hat{\mathcal{A}} \setminus \mathcal{S}).
\end{equation}

\subsection{Future Discussion} 
Rather than relying on known training datasets, which may only generalize to specific task domains, for topological structure optimization. Future work should explore task-oriented dynamic topology synthesis that can adapt during execution to unseen test instances or evolving task requirements. In addition, training-free approaches should also be explored in the future to construct communication structures on prior knowledge, such as agent roles or prompt-level heuristics.

\section{Use of Large Language Models}
We used ChatGPT solely to refine the manuscript’s language--improving grammar, phrasing, and clarity. All technical content, experimental design, analyses, and conceptual contributions were authored and verified by the authors.



\end{document}